\documentclass[twocolumn]{article}
\usepackage{amsfonts}
\usepackage{epsfig}
\usepackage{dcolumn}
\begin{document}
\title{Exploring Urban Environments \\ By Random Walks}

\author{Ph. Blanchard, D. Volchenkov
\vspace{0.5cm}
\\
{\small \it Bielefeld-Bonn Stochastic Research Center (BiBoS) },\\
{\small \it Bielefeld University, Postfach 100131, 33501 Bielefeld, Germany}\\
{\small \it Email: volchenk@physik.uni-bielefeld.de}
}
\date{\today}

\twocolumn[
\maketitle
\vspace{0.3cm}

\begin{flushright}
\textsl{To Ludwig Streit on his $70^{th}$ birthday with warm wishes}
\end{flushright}

\vspace{0.3cm}

\begin{abstract}
A complex web of roads, walkways and public transport systems can hide
areas of geographical isolation very difficult to analyze.
  Random walks are used to spot the structural details of urban fabric.
\end{abstract}

\vspace{0.2cm}

\leftline{PACS: 89.65.Lm, 89.75.Fb, 05.40.Fb, 02.10.Ox }
\vspace{0.2cm}

\leftline{Keywords: Space syntax, Queuing networks, Random walks}

\vspace{0.5cm}
]
\tableofcontents
\listoffigures
\newpage

\section{Introduction}
\label{sec:Introduction}
\noindent

In the present paper, we discuss the application of Markov processes to urban studies.
However, it is important to emphasize
 that the approach related to Markov processes
(such as random walks) can be used to analyze any type of complex networks.
The main technical
 idea beyond the method is to investigate the properties of networks by
  exploiting the spectral properties of self-adjoint operators (symmetric
  matrices) defined on their graph representations.
It is then obvious that
 a network can be effectively
investigated by spectral methods,
if it is not too large.
The spatial networks of human settlements give good examples -
for instance, the city of Paris (inside the Periphery Boulevard)
contains only 5131 interconnected open spaces.

A  city converts a space pattern into a pattern
of relationships. While discussing the impact a spatial structure has on
 the social establishment, it is inevitable to mention two Englishmen.
After
the British House of Commons
had been destroyed in 1941,
a long debate commenced in the Parliament
on whether the old House insufficient to
seat for all of its members be
rebuilt or a new modern House should be constructed.
In his speech to the meeting in the House
of Lords, October 28, 1943,  Sir Winston Churchill exclaimed, "We shape
our buildings, and afterwards our
buildings shape us" \cite{Churchill}, as if he believed that the movement
 pattern emerged in the old House of Common is responsible for the
 establishing and sustain of the British political tradition.  This
  intuitive knowledge had been conceived by another British gentleman,
   Bill Hillier,  professor at University College of London,
   within the concept of
 space syntax, a theory
developed in the late 1970s,
that seeks to reveal
the mutual effects of complex
spatial urban
networks on society and vice
versa \cite{Hillerhanson,Hillier2004}.
 The joint use of scarce space by many people creates life in cities which is driven largely
 by micro-economic factors which are invariant over various cultures that tends
to give cities similar structures.
Hereby, the street configuration itself naturally creates differential
patterns of occupancy (as an emergent phenomenon),
whereby some streets become, over time, more highly
used than others \cite{Iida}.
At the same time, a background residential space process driven by cultural
factors tend to make cities different from each other, so that
the emergent urban grid pattern containing an imprint of certain geometrical
 regularities forms a network of interconnected open spaces, being a
historical record of a city creating process driven by human activity and
containing traces of society and history \cite{Hillerhanson}.

 It is important to note that from the very beginning, graph theory had been
  developed in a close relation to urban studies:
 Leonard Euler's founding  paper
 on the Seven Bridges of
K\"{o}nigsberg
 published in 1736 is regarded as the first paper in the history of   graph theory,
  \cite{GraphTheory}.
In the present paper, we review the spectral algorithms designed
to analyze transport networks.
In particular, the method proves its efficiency
for revealing the disadvantageous areas designed to foster
crime, deprivation, and ghettoisation, \cite{Chown}.

In the forthcoming sections, we explain that city space syntax can
be considered as a queuing network operating in the free flow
regime (see Sec.~\ref{sec:Dual}). The steady state of a queuing
network (see Sec.~\ref{subsec:time}) is determined by the set of
automorphisms of the graph spanning the network
(Sec.~\ref{subsec:Automorphisms}). We shall show that these
automorphisms naturally correspond to random walks of the nearest
neighboring type, in which a traveller, after passing through an
open space, randomly chooses
 another space to move in
among those intersecting with the given one. How many
possibilities to move the walker has while in an arbitrary place?
- Indeed, it depends upon the type of the city he dwells in. We
have analyzed
 two German medieval organic cities, the city canal network in Venice, and the regular
street grid in Manhattan and found that while the organic towns are
relatively compact
(the majority of streets in them are just three steps apart from each other), the
bigger urban patterns demonstrate the structure of small worlds, in which a
 dense network of relatively short local streets is supported
by the long axial itineraries connecting the far-away districts
within the city that  essentially simplifies the navigation and
transportation tasks in the urban environment (see
Sec.~\ref{sec:Portraits}). Being a representation of the set of graph
automorphisms, random walks provide us with an effective tool for
describing the graph structure in details (see
Sec.~\ref{sec:Learn},~\ref{sec:One}). In Sec.~\ref{sec:Examples},
we illustrate the method by studying two graphs by random walks.
The first graph is the regular Petersen graph which contains just
10 nodes, and we investigate the network of Venetian 96 canals as
the second example.

The term "random walk" was originally proposed by K. Pearson in
1905 in his letter to {\it Nature} devoted to a simple model
describing a mosquito infestation in a forest: at each time step,
a single mosquito moves a fixed length, at a randomly chosen angle
(see \cite{Hugues}). Nowadays it is well known that random walks
could be effectively used in order to investigate and characterize
how effectively the nodes and edges of large networks can be
covered by different strategies,
 \cite{Tadic,Yang,Costa}.
Random walks help us to explore the structural properties of
complex urban networks which cannot be detected by the usual
 methods designed primarily to unravel the
structures
 of large quasi-random graphs.
 We conclude in the last section
(Sec.\ref{sec:Discussion}).

\section{City space syntax as a queueing network with a steady-state behavior}
\label{sec:Dual}
\noindent

Any graph representation  naturally arises
as the outcome of a categorization,
when we abstract a real world system by eliminating all but one of its features
 and by  grouping together items (or places) sharing a common attribute.
For instance, the common attribute of all open spaces
 in city space syntax is that we can move through them.
All elements called
 nodes that fall into one and the same set $V$ are considered as essentially
 identical; permutations of them within $V$ are of no consequence.
The symmetric group $\mathbb{S}_{N}$ consisting of all permutations of $N$
elements
($N$ being the cardinality of the set $V$) constitute therefore the symmetry group of $V$.
If we denote by $E\subseteq V\times V$ the set of ordered pairs of nodes called
edges, then  a graph is a map $G(V,E): E \to K\subseteq\mathbb{R}_{\,+}$.
We suppose that the graph is undirected and has no multiple edges. If two nodes are adjacent,
 $(i,j)\in E$, we write $i\sim j$. The degree of the node $i\in V$ is
the number of its neighbors in $G$,  $\deg_G(i)=k_i$.

The set of interconnected urban open spaces in which a pedestrian
or a vehicle, upon departures from one space, can join another
constitutes a queueing network that can be regarded as a graph
whose nodes represent the open spaces (streets, squares, and
round-abouts), and whose edges represent links between nodes
(junctions and crossroads). In the queueing theory
\cite{Queueing}, the paths along which a traveller may move from
 service station to service station are determined by the routing probabilities,
and then the theory of Markov chains provides the statistical models
 for the analysis of
queueing networks.
Travellers arriving to an open space are either
moving through it immediately or queuing until the space becomes available.
Once the space is passed through, the traveller is routed to its
next station, which is chosen according to a probability distribution
among all other open spaces linked
to the given one in the city.
However, if the destination space has
finite capacity then it may be full and then the traveller will be
blocked at its current location until the next space becomes available.


\subsection{Steady states in queuing networks}
\label{subsec:time}
\noindent

If the network of $N$ open spaces is not at all congested
(as long as inflows are compatible with system capacities), there is a network flow - say,
cars are travelling through it at some constant speed $v>0$ in the free flow regime.
Given $\left(X_t: t\geq 0\right)$, a i.i.d. random process characterized by
 some distribution
$f$ defined on $V$ and let $p_i >0$ be the arrival probability at
the node $i\in V$,  in a sequence of $n$ independent Bernoulli trials,
then travellers visit the node $i$ in average $np_i$ times.
If trials are conducted in a finite time interval $t$,
during which events occur in sequence, we denote
the proportion of time spent by a traveller in $i$ by $\lambda_i>0,$
\begin{equation}
\label{time_def}
 np_i\,=\,\lambda_i t,
\end{equation}
The above formula expresses the linear Ohm's law relating the time variable
with the number of independent Bernoulli trials, $t=(p_i/\lambda_i)n$.

Assuming the Kirchoff's current law, i.e.
the total flow into any node equals the flow out, then the proportions
 $\lambda_i$
can be obtained as the solutions of the global balance equations
(transport equations) satisfying a normalization condition,
\begin{equation}
\label{denisty}
\sum_{i\in V}\, \lambda_i\,=\,1.
\end{equation}
In the queuing network load in the stationary regime with free flow,
the  steady states (or densities) $\lambda_i>0$
can be interpreted
as the length distances of the correspondent streets $i\in V$
normalized by (\ref{denisty}).

In relation to a given distribution of $\lambda_i$, there are
 several
"performance indicators" characterizing the transport properties of
the queuing network
 with respect to the random process $\left(X_t: t\geq 0\right)$.
Given the arrival probability $p_i=t\lambda_i/n$,
no matter from which node $X_{t=0}=j$ the process starts,
the first arrival probability (FAP)
 that a traveller arrives at the node  $i\in V$ by time $t'> t$
 is given by
$
P\left(\left. t'>t\, \right|
X_{t'}=i \right)\,=\,\left(1-t\cdot\lambda_i/n\right)^n.
$
First encounter properties estimated by random walks
 play a crucial role in explorations of various
real-world networks, including epidemic spreading, transport in
disordered media, and neuron firing dynamics (see \cite{Condamin}
 and references therein).
In the limit of an infinite number of trials $n\to\infty$,
the FAP is asymptotically exponential,
\begin{equation}
\label{geom_limit}
\lim_{n\to\infty}P\left(\left. t'>t\, \right|
X_{t'}=i \right)\,
= \,  e^{-\lambda_i\,t},
\end{equation}
provided $\left(X_t: t\geq 0\right)$ preserves the proportion of
sojourn time $\lambda_i$ given by (\ref{time_def}). An exponential
distribution arises naturally when modelling the
 time between independent events that happen at a constant average rate.
The typical recurrence time which characterizes the random
duration elapsed between two consequent arrivals of the
traveller at $i\in V$
 in the free flow regime
described by the process $\left(X_t: t\geq 0\right)$
 is accounted by the probability density function (pdf)
$F_i(t) = \lambda_i e^{-\lambda_i\,t}. $
Then the expected
first-passage time (FPT) to the node $i\in V$ from a node randomly
chosen among all nodes in the networks is given by
\begin{equation}
\label{FPT}
\frac 1{\lambda_i}\,=\, \int_0^{\infty}t\, F_i(t)\, dt
\end{equation}
In complex networks, the value of FPT
crucially depends on the confining environment
and is not determined by the simple relation (\ref{FPT}).
 The FPT  has been recently recognized as a key quantity to specifying
 the transport-limited kinetics in many models
of transport in complex media,
\cite{Condamin}.

Other characteristic time can be used in order  to quantify the
long-run properties of the entire network. Let $\mathfrak{T}=\min\left(t:\bigcup_{\tau=0}^t
\{X_\tau\}=V\right)$ be
the covering time (also: the coupon-collector's time) required the
traveller visits all nodes of the closed queuing network (say,
rides along all streets in the city). The value of
$\mathfrak{T}$ crucially depends upon the local transport
properties of the network, \cite{JTP}. It
may be that
$\mathfrak{T}=\infty$ if the graph spanning the network is
directed, and
 the process $\left(X_t: t\geq 0\right)$ defined on that is irreversible.
The typical value of $\mathfrak{T}$ for a finite queuing network is
given by the unique solution $\tau$ of the equation
\begin{equation}
\label{Aldous}
\sum_{i\in V}\,e^{ -\lambda_i\,\tau}\,=\,1,
\end{equation}
expressing the idea that
after the tour the traveller inevitably returns into
one of the previously visited nodes \cite{JTP}.
It is easy to demonstrate that the
 minimal typical value, $\tau_{\min}=N\log N$, is achieved for the uniform distribution
$\lambda_i=N^{-1}.$
It is also clear that   $\tau\to\infty$
if the network contains least frequently visited nodes for which
 $\lambda_i\to 0.$

However, in many situations, the arrival probabilities $p_i$ are not
 independent.
It is intuitively clear that in real transport networks,
the arrival rates $p_i$ at the different
nodes  are correlated in accordance to the chance to
be joined by a path in the graph and therefore
may be sensitive to the particular graph symmetries.
In the forthcoming subsection, we derive
the distribution $\lambda_i$
associated to the set of automorphisms of
a simple graph $G$.

\subsection{Graph automorphisms and steady states}
\label{subsec:Automorphisms}
 \noindent

 Among all measures which can be
defined on the set of nodes $V$, the set of normalized measures
(densities) $\lambda_i\geq 0$, $i\in V,$ satisfying (\ref{denisty}),
are of essential interest since they express the conservation of a
quantity and therefore may be associated to some physical process.

The fundamental physical process defined on the graph $G(V,E)$ is
generated by the subset of its
automorphisms preserving the degrees of all nodes, $k_i$.

For each graph $G(V,E)$, there  exists a unique,
up to permutations of rows and columns,
adjacency matrix $\mathbf{A}$
identified with a linear endomorphism of
the vector space of all functions from $V$ into $\mathbb{R}$.
In the special case of a finite simple graph (an undirected graph
that has no self-loops), the adjacency matrix is a $(0,1)$-matrix
such that $A_{ij}=1$ if $i\ne j$,
  $i\sim j$, and $A_{ij}=0$ otherwise.
The degree of a node $i\in V$ is therefore given  by
\begin{equation}
\label{condition}
k_i\,=\,\sum_{i\sim j}\,A_{ij}.
\end{equation}
The set of  graph automorphisms, the
mappings of the graph to itself  preserving all of its structure,
 is
specified by the symmetric group $\mathbb{S}_N$
including all admissible permutations $\Pi\in \mathbb{S}_N$
taking the node $i\in V$ to $\Pi(i)\in V$.
 The representation of $\mathbb{S}_N$ consists of all
  $N\times N$ matrices ${\bf \Pi}_{\Pi},$ such that
  $\left({\bf \Pi}_{\Pi}\right)_{i,\,\Pi(i)}=1$, and $\left({\bf \Pi}_{\Pi}\right)_{i,j}=0$
if $j\ne \Pi(i)$.
A linear transformation of the adjacency matrix
\begin{equation}
\label{lin_fun}
Z\left({\bf A}\right)_{ij}\,=
\,\sum_{\,s,l=1}^N\, \mathcal{F}_{ijsl}\,A_{sl}, \quad \mathcal{F}_{ijsl}\,\in \,\mathbb{R}
\end{equation}
is a graph automorphism if
\begin{equation}
\label{permut_invar}
{\bf \Pi}_{\Pi}^\top\, Z\left({\bf A}\right)\,{\bf \Pi}_{\Pi}\,=\,
Z\left({\bf \Pi}_{\Pi}^\top\,{\bf A}\,{\bf \Pi}_{\Pi}\right),
\end{equation}
for any $\Pi\in \mathbb{S}_N$.
The latter relation is satisfied if
the entries of the tensor $\mathcal{F}$ in (\ref{lin_fun})
meets the following symmetry property:
 \begin{equation}
\label{symmetry}
\mathcal{F}_{\Pi(i)\,\Pi(j)\,\Pi(s)\,\Pi(l)}\,=\,
\mathcal{F}_{ijsl},
\end{equation}
for any $\Pi\in \mathbb{S}_N$.  The action of permutations
 preserves the conjugate classes of index
partition structures, so that any appropriate tensor $\mathcal{F}$
satisfying (\ref{symmetry}) can be expressed as a linear
combination of the following tensors: $ \left\{1,
\delta_{ij},\delta_{is},\delta_{il},\delta_{js},\delta_{jl},\delta_{sl},
\delta_{ij}\delta_{js},\delta_{js}\delta_{sl},\delta_{sl}\delta_{li},
\delta_{li}\delta_{ij},\right.\\
\left.
\delta_{ij}\delta_{sl},\delta_{is}\delta_{jl},\delta_{il}\delta_{js},
\delta_{ij}\delta_{il}\delta_{is} \right\}. $

Given a simple
undirected graph $G$,  then by
 substituting the above tensors into (\ref{lin_fun}) and taking
 account on symmetries we conclude that any arbitrary linear permutation invariant
 function $Z\left({\bf A}\right)$ must be of the following form
\begin{equation}
\label{lin_fun2}
Z\left({\bf A}\right)_{ij}\,=\,a_1+\delta_{ij}\,\left(a_2+a_3k_j\right)+
a_4\,A{}_{ij},
\end{equation}
with  $k_j=\deg_G(j)$ and $a_{1,2,3,4}$ being arbitrary constants.

If we require that the
  linear function $Z$ preserves the notion of connectivity,
\begin{equation}
\label{conn_nodes}
k_i\,=\,\sum_{j\in V}\,Z\left({\bf A}\right)_{ij},
\end{equation}
it is clear that we should take $a_1=a_2=0$ (indeed, the
contributions $a_1N$ and $a_2$ are incompatible with
(\ref{conn_nodes})) and then obtain a relation for the remaining
constants, $1-a_3=a_4$. Introducing the new parameter $\beta\equiv
a_4>0$, we reformulate (\ref{lin_fun2}) as follows,
\begin{equation}
\label{lin_fun3}
Z\left({\bf A}\right)_{ij}\,=\, (1-\beta)\,\delta_{ij} k_j +
\beta\,A_{ij}.
\end{equation}
If we express (\ref{conn_nodes})
in the form of a probability conservation relation,
 then the function
$Z\left({\bf A}\right)$ acquires  a probabilistic interpretation.
Substituting (\ref{lin_fun3}) back into (\ref{conn_nodes}), we obtain
\begin{equation}
\label{property}
\begin{array}{lcl}
1 & = & k_i^{-1}\sum_{j\in V}\, Z\left({\bf A}\right)_{ij}\\
  & = & \sum_{j\in V}\, (1-\beta)\,\delta_{ij} +\beta\,\frac{A_{ij}}{k_i}\\
 & = & \sum_{j\in V}\, T^{(\beta)}_{ij}.
\end{array}
\end{equation}
The  operator $T^{(\beta)}_{ij}$
is nothing else as
 the generalized random
walk transition operator if $0<\beta\leq k^{-1}_{\max}$ where
$k_{\max}$ is the maximal degree in the graph $G$. In the
"lazy" random walks defined by $T^{(\beta)}_{ij}$, a random walker stays
in the initial vertex with probability $1-\beta$, while it moves
to another node randomly chosen among its nearest neighbors with
probability $\beta/k_i$. If we take $\beta=1$, then the operator
$T^{(\beta)}_{ij}$ describes the usual random walks
extensively investigated in the classical surveys \cite{Lovasz}-\cite{Aldous}.

Being defined on a connected undirected graph, the matrix $T^{(\beta)}_{ij}$
is a real positive stochastic matrix, and therefore, in accordance to the
 Perron-Frobenius theorem \cite{PerronFrobenius}, its maximal
 eigenvalue equals 1, and it is simple. A left eigenvector,
\begin{equation}
\label{pi}
\pi\,T^{(\beta)}\,=\,\pi,
\end{equation}
 associated with the maximal eigenvalue 1 has positive entries satisfying
 the normalization condition (\ref{denisty}) independently of the value $\beta$,
\begin{equation}
\label{stationary_pi}
\pi_i\,=\,\frac{k_i}{\sum_{i\in V}k_i}.
\end{equation}
 It is interpreted as the unique
 equilibrium state $\pi$ (the stationary distribution of random walks).
If the graph $G$ is not bipartite, any density
$\sigma$ ($\sigma_i>0$, $\sum_{i\in V} \sigma _i=1$)
asymptotically tends to the stationary
distribution under the action of the transition operator  $T^{(\beta)}_{ij}$,
\begin{equation}
\label{limit}
\pi\,=\,\lim_{t\to\infty}\,\sigma\, \left(T^{(\beta)}\right)^{\,t}.
\end{equation}
If $G$ is regular, then $\pi$ is uniform. For a non-regular graph $G$,
this property is replaced by time reversibility (the balance equation)
\begin{equation}
\label{balance}
\pi_iT^{(\beta)}_{ij}\,=\,\pi_jT^{(\beta)}_{ji}, \quad i,j \in V.
\end{equation}
The stationary distribution of random walks $\pi$ is the unique density
associated to the automorphisms of a simple undirected graph $G$.

\section{Small worlds of big cities}
\label{sec:Portraits}
 \noindent

The main focus of the space syntax study is on the
relative proximity (or accessibility) between different locations
and associating these distances with densities and
intensities of human activity which occur at different open spaces and
along the links which connect them \cite{Hansen59,Wilson70,Batty}.
The decomposition of the city space into
a complete set of intersecting open spaces characterized by the certain
 traffic  capacities
produces a spatial networks which we call the dual graph representation of a city.

While identifying the spaces of motion which play the role of nodes in the
dual graphs of the
compact city patterns bounded by natural geographical
limitations, in the present paper
we implement the street named oriented identification principle,
the same which we used in our previous study
\cite{cities:2007}. The street named approach has been previously
used by different authors, \cite{Cardillo}-\cite{Crucitti}.
We have  generalized their approach
in a way to
account the possible discontinuities of streets.
Namely,
we assign an individual
street identification code (ID) to each continuous part of a street
even if all of them share the same street name. Then the dual graph
of the urban pattern
is constructed by
mapping spaces of motion coded with the same ID into nodes of the dual
graph and intersections among each pair of individual spaces
into edges connecting the corresponding nodes of the dual
graph.

The graph theoretic distance (also: depth) $d_{ij}$
between two locations in the urban pattern
is the least number of  steps (or the elementary
navigation actions) required
to reach the node $i$ from the node $j$
in the dual graph representation of the city.
In particular,
if two continuous streets intersect,
then the distance between them
 equals one.
Given the dual graph representation of a city, we can compute
the distances between all pairs of its nodes.

The number of neighbors a node has at a distance $d$ in the dual graph
quantifies the relative structural importance of the
location  in the urban environment, \cite{Hillier2004}.
The statistics of far away neighbors can be presented in the form
of cumulative distribution functions $\mathfrak{P}_d(n)$
quantifying the probabilities that a
 node has at least $n$ neighbors
 at a distance $d$. Being
monotonous functions of $n$,
cumulative distributions reduce the
noise in the distribution tails, however the adjacent points on
their plots are not statistically independent.
 \begin{figure}[ht]
 \noindent
\begin{center}
\epsfig{file=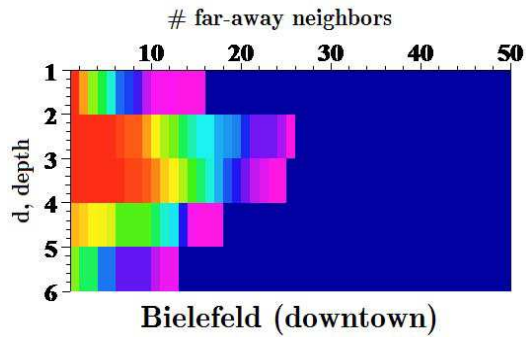,  angle= 0,width =7cm, height =4.5cm}
  \end{center}
\caption{\small The cumulative distributions of far-away neighbors
in the dual graph of Bielefeld downtown. The cumulative
degree distribution are shown in the first row. The cumulative
distributions of far-away neighbors are encoded in the second and
forthcoming rows. Probability is ranked from zero (dark blue) to 1
(red).}
\label{Fig1_21}
\end{figure}

\begin{figure}[ht]
 \noindent
\begin{center}
\epsfig{file=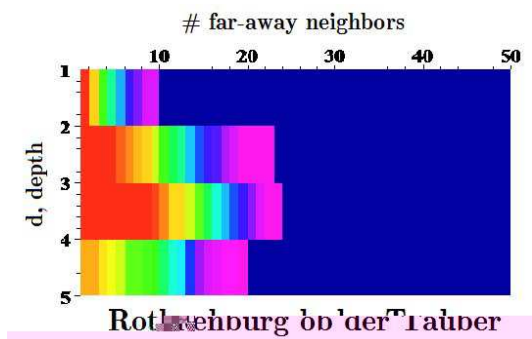,  angle= 0,width =7cm, height =4.5cm}
  \end{center}
\caption{\small  The cumulative distributions of far-away neighbors
in the dual graphs of Rothenburg ob der Tauber, Bavaria. The cumulative
degree distribution are shown in the first row. The cumulative
distributions of far-away neighbors are encoded in the second and
forthcoming rows. Probability is ranked from zero (dark blue) to 1
(red).}
\label{Fig1_22}
\end{figure}

For regular graphs, the number of far-away neighbors grows up
exponentially fast with the distance.
In Figs.~\ref{Fig1_21},~\ref{Fig1_22}, we have displayed the
cumulative distribution functions $P_d(n)$ quantifying the
probabilities
of that an average street is interconnected with at least $n$
other streets within a distance $d$
calculated
for  the dual
graphs of two German organic cities - the downtown of Bielefeld in
Westphalia and Rothenburg ob der Tauber in Bavaria. It is obvious
that  most of the streets in these towns can be reached in three
navigation steps. The compactness of German medieval towns
(see Figs.~\ref{Fig1_21},~\ref{Fig1_22}) uncovers the historical
functional background beyond their structure.

\begin{figure}[ht]
 \noindent
\begin{center}
\epsfig{file=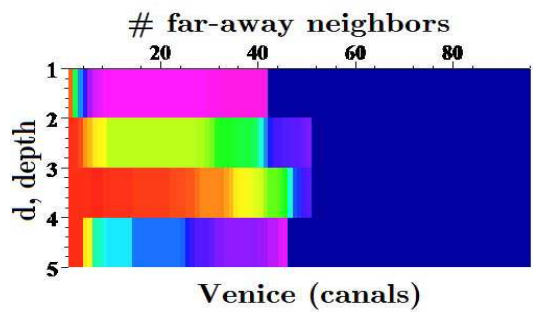,  angle= 0,width =7cm, height =4.5cm}
  \end{center}
\caption{\small The cumulative distributions of far-away neighbors
in the dual graph of Venetian canal network (96 canals).
Probability is ranked from zero (dark blue) to 1
(red).}
\label{Fig1_23}
\end{figure}

\begin{figure}[ht]
 \noindent
\begin{center}
\epsfig{file=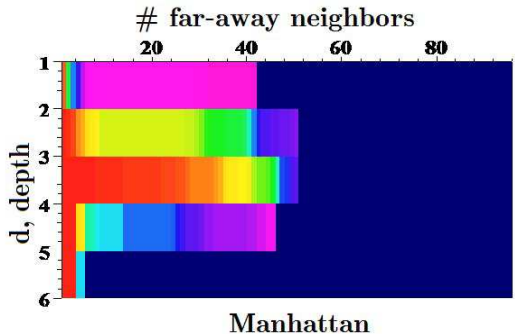,  angle= 0,width =7cm, height =4.5cm}
  \end{center}
\caption{\small The cumulative distributions of far-away neighbors
in the dual graph of the street grid in Manhattan.
Probability is ranked from zero (dark blue) to 1
(red).}
\label{Fig1_24}
\end{figure}

 The term of a {\it burgh} has been in use since the $12^\mathrm{th}$ century,
 when David I of Scotland  created the first Royal burghs.
Early burghs were granted the power to trade, which allowed them to control
 trade until the  $19^\mathrm{th}$ century,
\cite{Stewart}. Their spatial structure was shaped by the public
activities such as
 trade and exchange -
 ordered in such a way as to maximize the presence of
people in central areas.
In such an {\it organic} city, the majority
of streets are just by a few syntactic steps away from each
  other, so that the entire city is compact
that  can be clearly seen from  statistics of far away neighbors
 shown in Figs.~\ref{Fig1_21}-\ref{Fig1_24}.

It is interesting to mention that in the larger urban patterns such as the
 street grid in Manhattan and the canal network of Venice,
most of the nodes of spatial networks can still be reached in less as three
 navigation steps (see. Figs.~\ref{Fig1_23},\ref{Fig1_24}). However, there
 are also  noticeable fractions of relatively isolated places.
The structure of cumulative distributions shown in
Figs.~\ref{Fig1_23},~\ref{Fig1_24}
suggests
 that the correspondent
 graphs contain many  cliques (complete subgraphs),
 together with
subgraphs
being "almost" cliques.
In complex
network theory, this phenomenon is
called the {\it small
world} property.
 Small-world networks are characterized by a small diameter and a high clustering
coefficient having connections between almost any two nodes within
them. Hubs - nodes in the network
serving as the common connections mediating the short
path lengths between other edges are commonly associated with
small-world networks, \cite{Buchanan}.

The tendency to shorten distances in  urban
space networks induced by the public processes
 is complemented in the "small world" cities
(like those represented in Fig.~\ref{Fig1_23},~\ref{Fig1_24})
with the residential process  which shapes relations between
inhabitants and strangers preserving the original residential
culture against unsanctioned invasion of privacy. While
the majority
 of
streets and canals characterized with an excellent accessibility
promotes commercial activities and intensifies cultural exchanges,
 certain districts of such cities
demonstrate
 the opposite
tendency having residential areas relatively segregated
from the rest of the city, \cite{Hillier2004}.

The main technical problem of space syntax investigation
is to rank out the
open spaces in the city in accordance to their accessibility.

\section{From random walks to Euclidean spaces}
\label{sec:Learn}
\noindent

The idea of using
 the spectral properties of self-adjoint operators in order
to extract information about  graphs is standard in spectral graph theory
\cite{Chung:1997} and in theory of random walks on graphs \cite{Lovasz}-\cite{Aldous}.
In the following calculations, in the transition operator (\ref{property})
we take $\beta=1$
that allows to  compare  directly the forthcoming formulas
with those known from the classical surveys on random walks defined on graphs,
\cite{Lovasz}-\cite{Aldous}.

Among all measures which can be defined on the set of nodes $V$,
there is one associated to the stationary distribution of random
walks, $m=\sum_{i\in V}\pi_i\delta_i$, in which  $\delta_i$
is the vector of the canonical basis that
equals 1 at $i\in V$ and zero otherwise,
 with respect to which the
transition operator ((\ref{property}) with $\beta=1$) is self-adjoint
in the Hilbert space $\mathcal{H}(V)$,
\begin{equation}
\label{self_adj}
\widehat{T}\,=\,\frac 12\left( \pi^{1/2}T\pi^{-1/2}+\pi^{-1/2}T^{\top}\pi^{1/2}\right).
\end{equation}
In the above equation, $T^\top$ is the adjoint operator, and $\pi$ is considered as
the diagonal matrix
$\mathrm{diag}\left(\pi_1,\ldots,\pi_N\right)$. In particular, the elements of the
symmetric transition operator
defined on simple undirected graphs
equal
\begin{equation}
\label{T}
\widehat{T}_{ij}\, =\, \frac 1{\sqrt{k_ik_j}},\quad  \mathrm{iff}\,\, i\sim j,
\end{equation}
and zero otherwise.

The orthonormal ordered set of real
 eigenvectors $\psi_i$, $i=1\ldots N$, of the symmetric transition operator
 $\widehat{T}$
forms a basis in Hilbert space $\mathcal{H}(V)$.
The components of the first eigenvector
    $\psi_1$ belonging to the largest eigenvalue $\mu_1=1$,
\begin{equation}
\label{psi_1}
\psi_1
\,\widehat{ T}\, =\,
\psi_1,
\quad \psi_{1,i}^2\,=\,\pi_i,
\end{equation}
describes the connectivity of nodes ($\pi_i\propto k_i$).
The Euclidean norm of
the orthogonal complement of $\psi_1$,
 $\sum_{s=2}^N\psi_{s,i}^2=1-\pi_i$,
quantifies the probability that a random walker is not
in $i$.
The  eigenvectors,
 $\left\{\,\psi_s\,\right\}_{s=2}^N$, belonging to the eigenvalues
  $1>\mu_2\geq\ldots\mu_N\geq -1$
 describe the connectedness of the graph
$G$.
The exterior products of vectors calculated by means of determinants  are
used in Euclidean geometry to study areas,
volumes, and their higher-dimensional analogs, \cite{Algebra}.
Denote by $C_m$  the set of all subsets of $V$ containing  precisely
$1\leq m\leq N$ nodes. It is obvious that the size $\left|C_m\right|$ is given by the binomial coefficient,
$$
\left|C_m\right|\,=\, \frac{N!}{m!(N-m)!}.
$$
Let us note that
the eigenvectors $\psi_i$, $i=1\ldots N$, induce basis
 systems
for the spaces of all square summable $m$-point functions
$f:\left\{ i_1,\ldots i_m\in V\right\} \to \mathbb{R}$.
We consider
the matrix of eigenvectors of the symmetric transition
operator (\ref{psi_1}),
\begin{equation}
\label{determinant}
\Psi\,
= \,\left(
\begin{array}{ccc}
\psi_{1,1} &  \ldots & \psi_{1,N}\\
 \vdots&\vdots&\vdots\\
\psi_{N,1} &  \ldots & \psi_{N,N}
\end{array}
\right),
\end{equation}
in which the rows
 $\left\{\psi_k\right\}_{k=1}^N$ are orthonormal,
$\left(\psi_i,\psi_j\right)=\delta_{ij}$.
It is well known
that $\det \Psi =\pm 1$ and numerically equal to the content
of the parallelotope spanned by the unit eigenvectors $\left\{\psi_k\right\}_{k=1}^N$.

The nice property of minors
$\Psi^{i_1,\ldots i_m}_{s_1,\ldots s_m}$
 of order $m$ obtained by the deletion of $m$ rows $\left\{s_l\right\}$,
$l=1,\ldots m$,
and columns $\left\{ i_l\right\}$, $l=1,\ldots m$,
 from  $\det \Psi$ is that they are equal to their
algebraic complements in $\Psi$,
 multiplied by $\pm 1$, \cite{Muir:1960}.
 Moreover, the complete set of minors of order $m$
forms
an orthonormal system of $m-$point antisymmetric  functions,
\begin{equation}
\label{minors}
\sum_{i_1,\ldots i_m}\Psi^{i_1,\ldots i_m}_{s_1,\ldots s_m}
\Psi^{i_1,\ldots i_m}_{s'_1,\ldots s'_m}\,=\,
\delta_{s_1s'_1}\ldots \delta_{s_ms'_m},
\end{equation}
and can therefore be chosen as the basis in the Euclidean space
$\mathbb{R}^{\left|C_m\right|}$, \cite{Muir:1960}. In the next
section, we use the minors of order one to construct the embedding
of undirected graphs into Euclidean space. In particular, we
demonstrate that since any density $\sigma$ taken as an initial
distribution of random walkers
 on an
undirected graph
converges to the unique
 stationary distribution $\pi$ as $t\to \infty$,
this Euclidean space is $\mathbb{R}^{N-1}$.

The orthonormal systems of functions (\ref{minors}) are useful for
decomposing the data when analyzing the traffic properties of
transport networks. The data (say, of the real-time
 traffic flows and speed measurements)
from $m$ different  detectors  taken at different time slices
can be uniquely decomposed using the
orthonormal system (\ref{minors}) into smaller,
more manageable parts and then
compared allowing therefore an effective
classification of traffic speed patterns.

\section{Euclidean embedding \\ of graphs  by random walks}
\label{sec:One}
\noindent

In the present section, we consider the
orthogonal system of
 minors of order one,  $\Psi_{s}^{i}$, which are numerically
equal to the correspondent components of the eigenvectors
$\psi_{s,i}$. In particular, it is clear that the minors
$\left(\Psi_{1}^i \right)^2_i=\pi_i$ in accordance to
(\ref{psi_1}). The minors of order one ($m=1$) are associated to
the nodes of the graph and provide the basis for the Hilbert space
$\mathcal{H}(V)$. We demonstrate that a simple undirected graph
can be embedded into the $(N-1)$ Euclidean space, so that for any
pair of nodes the distance and angle can be uniquely defined.

It is important to note that Markov's symmetric transition operator
 $\widehat{T}$  defines a projection
 of any density $\sigma\in \mathcal{H}(V)$ on the eigenvector $\psi_1$
related to the
  stationary distribution $\pi$,
\begin{equation}
\label{project}
\sigma\,\widehat{T}\,
=\,\psi_1 + \sigma_\bot\,\widehat{T},\quad \sigma_\bot\,=\,\sigma-\psi_1,
\end{equation}
in which $\sigma_{\bot}$ is the vector belonging to the orthogonal complement of
$\psi_1$.
It follows from (\ref{limit}) that $\lim_{t\to\infty}\sigma_\bot\widehat{T}^{t}=0,$
so that the vector  $\sigma_{\bot}$ characterizes the transient
process induced by $\sigma$.

In space syntax, we are interested in a comparison between the
different densities defined on the nodes of the graph.
 It is convenient
to compare them in accordance to their proximity
to the unique stationary
distribution of random walks $\pi$ defined on the graph $G$.
Since all components $\psi_{1,i}>0$,
we can rescale the density $\sigma$ by dividing its
components by the components of $\psi_1$,
\begin{equation}
\label{rescaling}
\widetilde{\sigma_i}\, =\,\frac{\sigma_i}{\psi_{1,i}}
\,=\, \frac{\sigma_i}{\sqrt{\pi_i}}.
\end{equation}
It is clear that any two rescaled densities
$\widetilde{\sigma},\widetilde{\rho}\,\in\,\mathcal{H}$ differ
 with respect to random walks only by their dynamical components
orthogonal to $\psi_1$,
$$
\left(\widetilde{\sigma}-\widetilde{\rho}\right)\,
 \widehat{T}^t\,=\,\left(\widetilde{\sigma}_\bot -\widetilde{\rho}_\bot\right)\,
\widehat{T}^t,
$$
 for all $t>0$.

Therefore, we can define the
distance  $\|\ldots\|_T$ based on random walks between any two densities  by
\begin{equation}
\label{distance}
\left\|\,\sigma-\rho\,\right\|^2_T\, =
\, \sum_{t\,\geq\, 0}\, \left\langle\, \widetilde{\sigma}_\bot -\widetilde{\rho}_\bot\,\left|\,\widehat{T}^t\,
\right|\, \widetilde{\sigma}_\bot -\widetilde{\rho}_\bot\,\right\rangle.
\end{equation}
 or, using the spectral
representation of $\widehat{T}$,
\begin{equation}
\label{spectral_dist}
\begin{array}{l}
\left\|\sigma-\rho\right\|^2_T
 \\
=
\sum_{t\,\geq 0} \sum_{s=2}^N\, \mu^t_s \left\langle
\widetilde{\sigma}^\bot -\widetilde{\rho}^\bot|\psi_s\right\rangle\!\left\langle \psi_s
| \widetilde{\sigma}^\bot -\widetilde{\rho}^\bot\right\rangle
\\
=
\sum_{s=2}^N\,\frac{\left\langle\, \widetilde{\sigma}_\bot -\widetilde{\rho}_\bot\,|
\, \psi_s\,\right\rangle\!\left\langle\, \psi_s\,
| \,\widetilde{\sigma}_\bot -\widetilde{\rho}_\bot\,\right\rangle}{\,1\,-\,\mu_s\,},
\end{array}
\end{equation}
where we have used  Dirac's bra-ket notations especially
convenient for working with inner products and
rank-one
operators in Hilbert space.

If we introduce a new inner product for
densities $\sigma,\rho \in\mathcal{H}(V)$
by
\begin{equation}
\label{inner-product}
\left(\,\sigma,\rho\,\right)_{T}
\,= \,  \sum_{s=2}^N
\,\frac{\,\left\langle\,  \widetilde{\sigma}_\bot\,|\,\psi_s\,\right\rangle\!
\left\langle\,\psi_s\,|\, \widetilde{\rho}_\bot \right\rangle}{\,1\,-\,\mu_s\,},
\end{equation}
then (\ref{spectral_dist}) is nothing else but
\begin{equation}
\label{spectr-dist2}
\left\|\,\sigma-\rho\,\right\|^2_T\, =
\left\|\,\sigma\,\right\|^2_T +
\left\|\,\rho\,\right\|^2_T  -
2 \left(\,\sigma,\rho\,\right)_T,
\end{equation}
 where
\begin{equation}
\label{sqaured_norm}
\left\|\, \sigma\,\right\|^2_T\,=\,
\,\sum_{s=2}^N \,\frac{\left\langle\,  \widetilde{\sigma}_\bot\,|\,\psi_s\,\right\rangle\!
\left\langle\,\psi_s\,|\, \widetilde{\sigma}_\bot\, \right\rangle}{\,1\,-\,\mu_s\,}
\end{equation}
is the square
of the norm of  $\sigma\,\in\, \mathcal{H}(V)$ with respect to
random walks defined on the graph $G$.

We conclude the description of the $(N-1)$-dimensional Euclidean
space structure of $G$ induced by
  random walks by mentioning that
given two densities $\sigma,\rho\,\in\, \mathcal{H}(V),$ the
angle between them can be introduced in the standard way,
\begin{equation}
\label{angle}
\cos \,\angle \left(\rho,\sigma\right)=
\frac{\,\left(\,\sigma,\rho\,\right)_T\,}
{\left\|\,\sigma\,\right\|_T\,\left\|\,\rho\,\right\|_T}.
\end{equation}
Random walks embed connected undirected graphs into the Euclidean
space $\mathbb{R}^{N-1}$. This embedding  can be used in order to
compare the accessibility properties of individual
nodes and to construct
 the optimal coarse-graining
representations of the transport networks.

Indeed, the vector $\delta_i$ of the canonical basis is the particular
case of density, which is zero for all nodes excepting for
the node $i\in V$ where it equals 1.
In accordance to (\ref{sqaured_norm}), the density $\delta_i$
acquires the norm $\left\|\,\delta_i\,\right\|_T$
associated to random walks defined on $G$.
In the theory of random walks \cite{Lovasz},
 its square,
\begin{equation}
\label{norm_node}
\left\|\,\delta_i\,\right\|_T^2\, =\,\frac 1{\pi_i}\,\sum_{s=2}^N\,
\frac{\,\psi^2_{s,i}\,}{\,1-\mu_s\,},
\end{equation}
 gets
a clear probabilistic interpretation
expressing the spectral formula of
 access time to the  node $i\in V$,
 the expected number
of  steps
required for a random walker
to reach the node
$i$ starting from an
arbitrary
node  chosen randomly
among all other
nodes of the graph $G$ with respect to
the stationary distribution of random walks $\pi$.
The squared  norm of the canonical vector $\left\|\,\delta_i\,\right\|_T^2$
can be used in order to characterize the access to the node $i$ from other
nodes of the graph $G$.
By the inequality between arithmetic and harmonic means, it is easy to prove \cite{Lovasz}
that
\begin{equation}
\label{norm_node_2}
\left\|\,\delta_i\,\right\|_T^2\, \geq\,\frac {\left(1-\pi_i\right)^2}{\pi_i},
\end{equation}
so that the nodes which are
difficult to reach ($\pi_i\ll 1$) are also isolated from the rest of the graph.

The Euclidean distance between any two nodes of the graph $G$ is
calculated as the distance (\ref{spectral_dist}) induced by random walks
between two
canonical vectors  $\delta_i$ and
$\delta_j$,
\begin{equation}
\label{commute}
K_{i,j}\,=\,\left\|\, \delta_i-\delta_j\,\right\|^2_T\,=\, \sum_{s=2}^N\,
\frac 1{1-\mu_s}\left(\frac{\psi_{s,i}}{\sqrt{\pi_i}}-\frac{\psi_{s,j}}{\sqrt{\pi_j}}\right)^2,
\end{equation}
gives the spectral representation
of
 commute time defined in the theory of random walks
on undirected graphs as  the expected number of steps required for a random
walker starting at $i\,\in\, V$ to visit $j\,\in\, V$ and then to
return back to $i$,  \cite{Lovasz}.

The commute time which plays the role of the Euclidean distance
between the nodes of the graph can be represented as a sum,
$K_{i,j}=H_{i,j}+H_{j,i}$, in which
\begin{equation}
\label{hitting}
H_{i,j}\,=\,\left\|\,\delta_i\,\right\|^2_T - \left(\,\delta_i,\delta_j\,\right)_T
\end{equation}
is the first hitting time which plays probably the
most important role in the quantitative theory of random walks \cite{Lovasz}.
It quantifies the expected number of steps
before node $j$ is visited if starting from node $i$.  The first
hitting time $H_{i,j}$ satisfies the equation
\begin{equation}
\label{eq2}
H_{i,j}\,=\, 1+ \frac 1{k_i}\sum_{l\sim i} H_{l,j}
\end{equation}
with the intimal condition $H_{i,i}=0$. The equation (\ref{eq2})
expresses the fact that the first step takes the random walker to
a neighbor $l \sim i$ and then he has
 to reach $j$ from there.

It is important to mention that
the cosine of an angle calculated in accordance to
 (\ref{angle}) has the structure of
Pearson's coefficient of linear correlations
 that reveals it's natural
statistical interpretation.
The notion of angle between any two nodes of the
graph arises naturally as soon as we
become interested in
the strength and direction of
a linear relationship between
two random variables,
the flows of random walks moving through them.
If the cosine of an angle (\ref{angle}) is 1
(zero angles),
there is an increasing linear relationship
between the flows of random walks through both nodes.
Otherwise, if it is close to -1,
  there is
a decreasing linear relationship.
The  correlation is 0
if the variables are linearly independent.
It is important to mention that
 as usual the correlation between nodes
does not necessary imply a direct causal
relationship (an immediate connection)
between them.

\section{Petersen graph and Venetian Canals}
\label{sec:Examples}
\noindent

\begin{figure}[ht]
 \noindent
\begin{center}
\epsfig{file=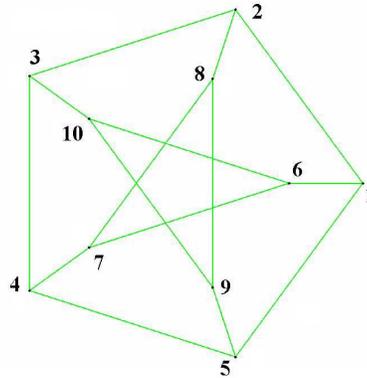,  angle= 0,width =5cm, height =5cm}
  \end{center}
\caption{\small The Petersen graph.}
\label{Fig1}
\end{figure}

\begin{figure}[ht]
 \noindent
\begin{center}
\epsfig{file=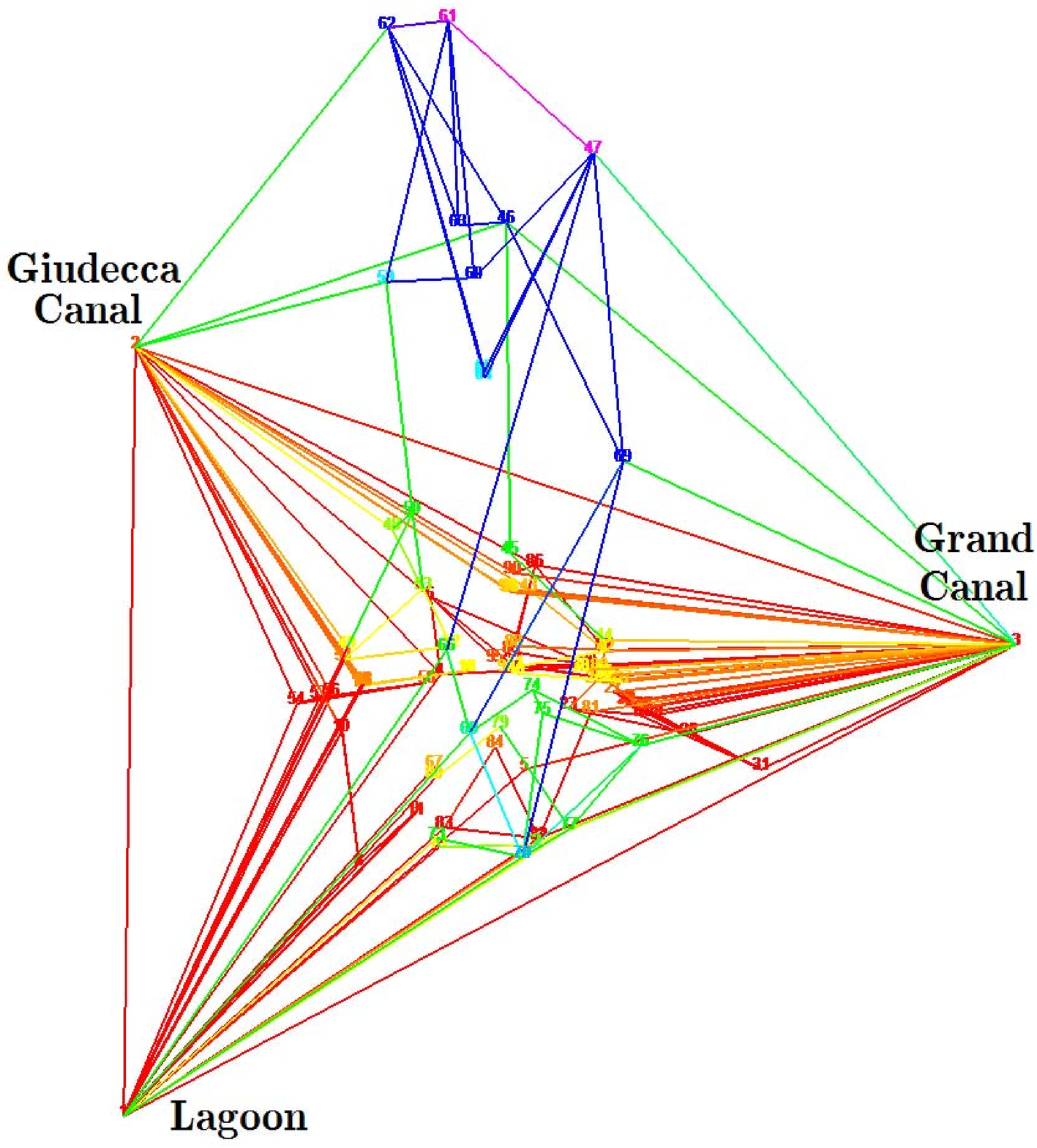,  angle= 0,width =7cm, height =7cm}
  \end{center}
\caption{\small The dual graph representation of the spatial network of 96 Venetian canals }
\label{Fig2}
\end{figure}

In the present section, we construct and investigate
 the Euclidean embedding of
two graphs. The first one we study is the Petersen graph with
10 nodes (see Fig.~\ref{Fig1}). Another example is the spatial
network of 96 Venetian canals which serve the function of roads in
the ancient city that stretches across 122 small islands (see
Fig.~\ref{Fig2}). While identifying canals over the plurality of
water routes on the city map of Venice, the canal-named approach
has been used, in which two different arcs of the city canal
network were assigned to the same identification number provided
they have the same name.
The Petersen graph is a regular graph, $k_i=3$, $i=1,\ldots 10$,
so that $\sum_{i}k_i=30$, and the stationary distribution of
random walks is uniform, $\pi^{(\mathrm{Pet})}_i=0.1$. The
spectrum of the random walk transition operator
(\ref{self_adj}) defined on the Petersen graph
contains the Perron eigenvalue $\mu_1=1$ which
is simple, then the eigenvalue $\mu_2=1/3$ with multiplicity $5$, and
$\mu_3=-2/3$ with multiplicity $4$. Therefore, there are just $3$ linearly
independent eigenvectors, and two eigensubspaces for which the
orthonormal basis vectors can be calculated, so that the resulting
matrix of eigenvectors and basis vectors which we use in
(\ref{norm_node}-\ref{commute}) always has full column dimension.
Random walks embed the Petersen graph into a $9$-dimensional
Euclidean space, in which all nodes have equal norm
(\ref{norm_node}), $\left\| i\right\|_T= 3.14642654$ meaning that
the expected number of  steps a random walker starting from a node
chosen randomly with probability $p=0.1$ reaches any node in the
Petersen graph equals $9.9$.
\begin{figure}[ht]
 \noindent
\begin{center}
\epsfig{file=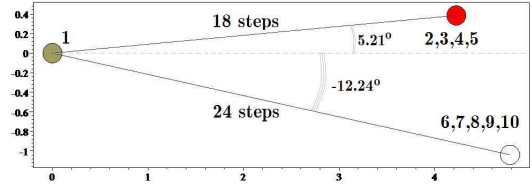,  angle= 0,width =7cm, height =2.5cm}
  \end{center}
\caption{\small The Euclidean space embedding of the Petersen
graph drawn with respect to the node $\# 1$.}
\label{Fig3}
\end{figure}
Indeed, the structure of $9$-dimensional vector space  induced by
random walks defined on the Petersen graph cannot be represented
visually, however if we choose one node as a point of reference, we
can draw its 2-dimensional projection by arranging other nodes at
the distances calculated accordingly to (\ref{commute}) and under
the angles (\ref{angle}) they are with respect to the
chosen reference node (see Fig.~\ref{Fig3}).

It is expected that a random walker starting at the node $1$
visits any peripheral node ($2,3,4,5$)
and then returns back in 18 random steps, while   24
random steps are expected in order to visit any internal node in the deep of the graph
($6,7,8,9,10$). It is also obvious that while the
linear relationship
between the  random walks flows
through the node $ 1$ and those through any peripheral node
is positive, it is negative  with respect to
 the flows passing through the internal nodes.
Due to the symmetry of the Petersen graph, the figure displayed on Fig.~\ref{Fig3}
is essentially the same if we draw it with respect to any other periphery node
 ($2,3,4,5$). It is also important to note that it turns to be mirror-reflected
if we draw it with respect to any internal node
($ 6,7,8,9,10$). Therefore, we can conclude that the
 Petersen graph contains two groups of nodes, at the
 periphery and in the depth, which appears to be
as much as a quarter more isolated
  from one another than the nodes within each group
 (18 random steps vs. 24 random steps).
It is clear that the $9-$dimensional embedding of the Petersen graph
into Euclidean space is characterized by the highest degree of symmetry.

\begin{figure}[ht]
 \noindent
\begin{center}
\epsfig{file=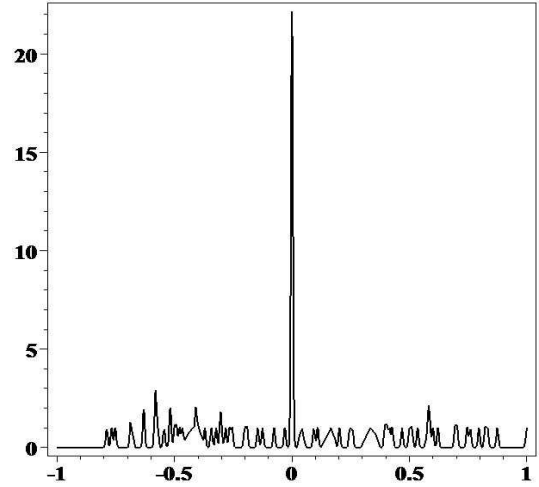,  angle= 0,width =7cm, height =6.5cm}
  \end{center}
\caption{\small The spectrum of the Markov transition operator
(\ref{T}) defined on the spatial network of Venetian canals.}
\label{Fig4}
\end{figure}

The dual graph of the spatial network of Venetian canals
(Fig~\ref{Fig2}) is much more complicated than the Petersen graph.
We construct it by mapping every canal into an individual node of
the dual graph and connecting the pairs of nodes by arcs when
these canals intersect. The resulting graph is far from being
regular, so that the stationary distribution of random walks
defined on it is not uniform. In \cite{cities:2007}, we have
discussed that it is not evident if the degree distributions
observed in
compact urban patterns and in the Venetian canal network, in particular, follow a
power law. The spectrum of the symmetric Markov transition operator
(\ref{self_adj})
  defined on that is presented in Fig.~\ref{Fig4}.
The transition matrix $\widehat{T}$ for the canal network in Venice
is strongly defective. In particular, it contains the eigenvalue
$\mu=0$ with multiplicity 22. This degeneracy indicates the presence
of a complete bipartite subgraph in the spatial network of
Venice shown in Fig.~\ref{Fig2}.
\begin{figure}[ht]
 \noindent
\begin{center}
\epsfig{file=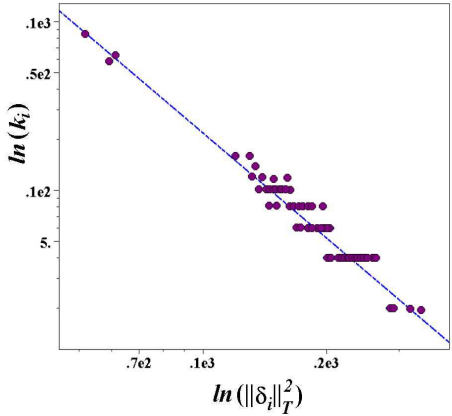,  angle= 0,width =6cm, height =5.5cm}
  \end{center}
\caption{\small The scatter plot (in the log-log scale)
of the connectivity vs. the value of access time to a node in
the dual graph representation of 96 Venetian canals.
Three data points characterized by the shortest access times represent the
main water routes of Venice: the Lagoon of Venice, the Giudecca canal, and
the Grand canal. Four data points of the worst accessibility are for the
canal subnetwork of Venetian Ghetto.
The slope of the regression
 line equals 2.07.}
\label{Fig5}
\end{figure}
In urban
spatial networks encoded by their dual graphs,
 the values of access time
(\ref{norm_node})
vary strongly from one canal
to another being very long for
topologically isolated places.
Three data points characterized by
the shortest access times shown in Fig.~\ref{Fig5} are associated to the Lagoon
of Venice, the Giudecca canal, and
the Grand canal - the most central water routes in the city canal network.
Four data points characterized by the worst accessibility represent the
canal subnetwork of Venetian Ghetto.

The values  of access times observed in
the dual graph  of Venetian canals
scale with the connectivity of canals:
the slope of the regression
 line equals 2.07.

\begin{figure}[ht]
 \noindent
\begin{center}
\epsfig{file=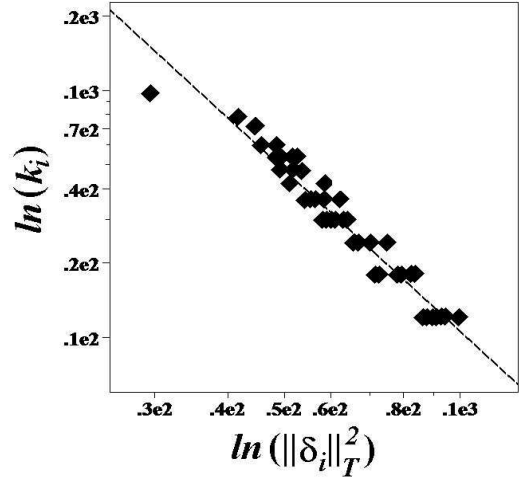,  angle= 0,width =7cm, height =6.5cm}
  \end{center}
\caption{\small The scatter plot of the connectivity vs. the norm a node in
the dual graph representation of 50 streets in the downtown of Bielefeld.
The slope of the regression
 line equals 2.17.}
\label{Fig5a}
\end{figure}
A similar power law can be observed for the isolation patterns in other
 urban environments. For example, in Fig.~\ref{Fig5a}, we have sketched
 the scatter plot of the connectivity vs. the norm a node in
the dual graph representation of 50 streets in the downtown of Bielefeld.
 Professor Dr. Ludwig Streit works at the University of Bielefeld since 1972.

\begin{figure}[ht]
 \noindent
\begin{center}
\epsfig{file=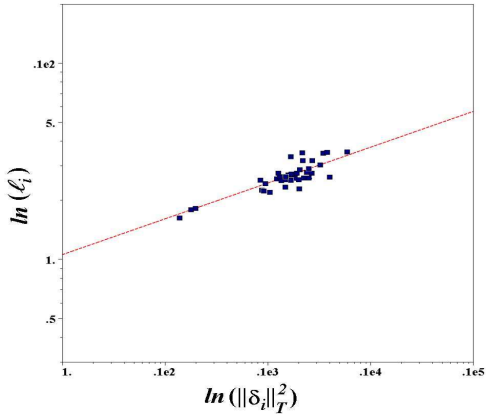,  angle= 0,width =6.5cm, height =5.5cm}
  \end{center}
\caption{\small The  scatter plot
(in the log-log scale)
of the  mean shortest distance vs. the value of access time
for the network of  Venetian canals. The plot indicates
a positive relation with the slope of regression line equals 0.18.}
\label{Fig5aa}
\end{figure}

In
 the space syntax analysis,
the mean shortest distance
 from a given node $i$ to any other node
in the  graph,
\begin{equation}
\label{meandepth}
\ell_i\,=\,\frac {1}{N-1}\sum_{j\in V}d_{ij},
\end{equation}
is used for
characterizing the level of integration of the
node in the city pattern,
\cite{Jiang98}.
The relation between the mean shortest distance
to a node $\ell_i$
and the value of access time to the same node
$\left\|\delta_i \right\|^2_T$ is very complicated and
strongly depends on the topology of the graph.
In Fig.~\ref{Fig5aa}, we have shown the  scatter plot
(in the log-log scale)
of the  mean shortest distance vs. the value of access time
for the network of  Venetian canals. The  plot Fig.~\ref{Fig5aa}
 indicates a positive relation between $\ell_i$
and $\left\|\delta_i \right\|^2_T$
with the slope of the regression line equals 0.18.

\begin{figure}[ht]
 \noindent
\begin{center}
\epsfig{file=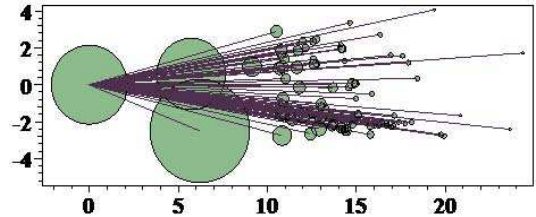,  angle= 0,width =7cm, height =3.5cm}
  \end{center}
\caption{\small The 2-dimensional projection of the 95-dimensional
 Euclidean spaces associated to random walks defined on the city
 canal network built from the perspective of the Grand canal of Venice
chosen as the origin. The labels of the
horizontal axes display the expected number of random walk steps.
The labels of the
vertical axes show the degree of nodes (radiuses of the disks). }
\label{Fig6}
\end{figure}

The 2-dimensional projection
of the  Euclidean space of 96 Venetian canals  set up by random walks
drawn for the
 the Grand Canal of Venice (the point $(0,0)$) is shown in Fig.~\ref{Fig6}.
Nodes of the dual graph representation of
 the canal network in Venice
are shown by disks with radiuses taken equal to the degrees of the
nodes. All distances between the chosen origin and other nodes of the
graph (Fig.~\ref{Fig2}) have been calculated in accordance to
(\ref{commute}) and (\ref{angle}) has been used in
order to compute angles between nodes.
Canals negatively correlated with the Grand Canal of Venice are
set under negative angles (below the horizontal), and under
positive angles (above the horizontal) if otherwise.
It is evident from Fig.~\ref{Fig6}
that disks of smaller radiuses
demonstrate a clear
tendency to be located far
away from the origin
being characterized by
 excessively long commute times with the reference node (the Grand canal of Venice),
while the large disks which stand in Fig.~\ref{Fig6} for the main water routes
are settled in the closest proximity to the
origin that intends an immediate access to them.

\section{Discussion and Conclusion}
\label{sec:Discussion}
\noindent

From Euler's
time, urban design and townscape studies were the sources of
inspiration for the network analysis and graph theory.
It is
common now that networks are the reality of urban renewals
\cite{renewal}. Flows of pedestrians and vehicles through a city
are dependent on one another and that requests for organizing them
in a network setting. Moreover, the networking is structurally contagious.
In order to be able to
 master a network
effectively,
an authority
 should also constitute
a network structure,
probably
as complicated as the one that
it supervises. A complex network of city itineraries
that we can experience in everyday life
appears as
the result of multiple
complex interactions
between various transport,
 social,
and economical networks.

In the present paper, we have developed a self-consistent approach
for exploring the structure of urban environments based on the use of Markov chains
(random walks)
defined on the dual graph  of the city. The stochastic transition matrix of
random walks gives a matrix representation for the group of automorphisms of the
graph $G(V,E)$.
The expected number of steps required to reach a node of the graph
 starting from a node chosen from the stationary distribution over $V$
given by (\ref{norm_node})
can be considered as a characteristic time scale
 $\mathcal{T}_{[i]}=\left\|\delta_i\right\|^2_T$ quantifying the
accessibility of the node $i$ in the graph $G$ by random walks.
The data of
scatter plots displayed in
Figs.~\ref{Fig5},\ref{Fig5a} show that in the studied urban patterns
 $\mathcal{T}_{[i]}$
scales with the degree of vertex, so that
$\mathcal{T}(k)\propto k^{-2-2\varepsilon}$, with some $\varepsilon>0$
($\varepsilon =0.085$ for the downtown of Bielefeld and
$\varepsilon=0.035$ for the canal network of Venice).
We have also  discussed in Sec.~\ref{subsec:time} that
in the free flow regime of a queuing network
preserving the proportion of time  $\lambda_i$ spent in $i\in V$
constant 
(which then can be considered as the normalized length distance of
the given street $i$), the expected first passage time to the node
from a node random chosen among all nodes of the queuing network
equals $\lambda_i^{-1}$. While equalizing these passage times, we
obtain that in the free flow regime  the proportion of time
$\lambda$ spent by a moving agent in an open place of the given
transport network
 scales with its connectivity as
\begin{equation}
\label{lambda_vs_k}
\lambda(k)\,\propto\,k^{2+2\varepsilon},
\end{equation}
where the value $\varepsilon>0$ is determined by the entire topology of the city.

It is well documented that isolation worsens
an area's economic prospects by reducing
opportunities for commerce, and engenders a feeling of isolation
for the inhabitants, both of which can fuel poverty and crime \cite{Hillier2004}.
The method we have introduced  in the present paper
 could easily be used to identify isolated neighborhoods
  in big cities with a complex web of roads, walkways and public transport systems
 spotting hidden areas of geographical isolation in the urban landscape.
The method can be used to analyze complex transport networks of
any type.

\section*{Acknowledgment}
\label{Acknowledgment}
\noindent

The support from the Volkswagen Foundation (Germany) in the
framework of the project "{\it Network formation rules, random set
graphs and generalized epidemic processes}" is gratefully
acknowledged.

We wish to thank Thomas K\"{u}chelmann for his
thorough reading of the paper and his excellent comments on that.

\end{document}